\documentclass[11pt]{article}
\usepackage{hyperref}
\begin{document}
 \renewcommand{\thefootnote}{*}
\title{Biological Observer-Participation and Wheeler's `Law without Law'
\footnote{Based on talk given at the ACIB '11 conference, INBIOSA project\cite{INBIOSA}, to be published in \textit{`Integral Biomathics: Tracing the Road to Reality', Proceedings of iBioMath 2011, Paris and ACIB '11, Stirling UK,} P. L. Simeonov, L. S. Smith, A. C. Ehresmann (Eds.), Springer-Verlag, 2012.}}
\author{Brian D. Josephson\\Trinity College, Cambridge CB2 1TQ, UK}
\date{October 2011}
\maketitle
\thispagestyle{empty}
\setcounter{page}{0}
\setcounter{secnumdepth}{0}
\section{Abstract}
It is argued that at a sufficiently deep level the conventional quantitative approach to the study of nature faces difficult problems, and that biological processes should be seen as more fundamental, in a way that can be elaborated on the basis of Peircean semiotics and Yardley's Circular Theory.  In such a world-view, Wheeler's \textit{observer-participation} and emergent law arise naturally, rather than having to be imposed artificially. This points the way to a deeper understanding of nature, where meaning has a fundamental role to play that is invisible to quantitative science.

\subsection{Keywords}
Observer-participation, Peirce, semiotics, signs, interpretation, emergence, complexity, cognitive development, space-time generation, wholeness, symmetry, Circular Theory.
\newpage
\setcounter{secnumdepth}{1}
\setcounter{section}{0}

\section{Introduction}
It is commonly assumed that nature can be described in terms of fixed mathematical laws.  However, the discovery that the Standard Model cannot be reconciled with general relativity in a straightforward way has created problems for this point of view. An alternative is Wheeler's proposal to the effect that participation by observers, as postulated in some formulations of quantum mechanics, is the mechanism whereby physical laws emerge.  According to Wheeler, that principle might suffice to build everything\cite{lawless}.

In Wheeler's article the gap between acts of observer-participancy and physical reality was not filled in, an insufficiency that we attribute to the absence of an appropriate theory of observation.  In the following we discuss a biologically oriented scheme where observation plays a central role, and show how it can lead to the emergence of physical laws.

 The structure of this scheme can be summarised as \\
\begin{math}
primordial\ reality \rightarrow circular\ mechanics \rightarrow semiotics\ and\ structure \rightarrow technological\ development \rightarrow regulatory\ mechanisms \rightarrow emergent\ laws.
\end{math}
Here `circular mechanics' is a reference to a generic scheme of biological organisation proposed by Yardley\cite{circular}, encompassing among its aspects sign processes in accord with the semiosis concepts of Peirce\cite{semiotics}, which in turn underlie processes of a technological character, among which we hypothesise are the capacity to form systems such as our universe, to which laws of a mathematical kind are applicable.  In this way, we are able to link life, viewed from a generic point of view, to the origin of universes.

We discuss first of all the relationship between idealised situations in physics which can be characterised precisely in mathematical terms on the one hand and on the other, biology, which it will be argued is primarily concerned with patterns and only secondarily with quantities. The characteristics of biosystems are then related to the forward-looking role of signs, and to circular theory approach, thus paving the way to a more detailed analysis of universe generation.
 
\section{Physics vs. biology; mathematics vs. semiosis}

Theoretical physics is mathematics-based, typically involving differential equations with respect to time.  Such a mathematical approach carries the presumption that systems found in nature can be represented adequately by explicit formulae.  Experimental biology gives the appearance of demonstrating the derivability of life from conventional physics, such investigations uncovering a great variety of processes that accord with known physics as well as having biological functions.  However, things are not what they seem.  To see this, compare life with a phenomenon of physics such as superconductivity.  In the latter case there is a specific model, the BCS model, defined by a specific mathematical expression, which accords well with many experimental observations.  Small changes in the model would have small consequences, and would not affect this agreement.  Biosystems differ in that fine details may drastically affect behaviour; rather than there being a specific model there is a \textit{landscape} of possibilities, with only the peaks reflecting viable systems.  Thus the properties of biosystems cannot be accounted for on the basis of a first-principles computation, which could not apply to such a landscape.
    
Biosystems must therefore be addressed in a way different from the way systems that are the subject of mathematical physics are normally studied. They can be conceived of as systems that have passed certain tests, a situation similar to that of prime numbers, where in general a number can be shown to be prime only by testing for factors, rather than there being a formula that generates all primes.   Despite the absence of such a formula, passing such tests has important implications.  The situation addressed by G\"odel, whereby there exist true statements that cannot be proved starting from specified axioms, is similar in the way it demonstrates limits of specifiability.  In the biosystem case, the test-passing factor is related to viability, and is also responsible for different instances of an organism behaving similarly, which permits their non-quantitative analysis.

\section{The forward-looking aspect of life; semiosis}

One way in which life differs from nature is general is the way it creates its own structures, in a way that does not admit of any very direct mathematical interpretation.  Rather, in life we find systems that have come into existence that are able to pass particular tests, as required for the survival of the given system.  One aspect of this is the \textit{semiosis} discussed by Peirce\cite{semiotics},  Semiotics empasises the role of information processing and more specifically the importance of the interpretation of \textit{signs}, in the light of the fact that at the cognitive level the appropriate use and interpretation of signs is essential.  In Peirce's scheme there is a specific, possibly context dependent, relationship between signs, and objects to which they are linked, with a third element, the \textit{interpretant}, having the role of linking them.  Typically, a complicated interpretant mechanism links the simpler sign and object, reliably producing a well defined situation linked to the sign.

The role played by signs in biological situations can be illustrated by the situation of road traffic.  The fact that cars collide with each other much less frequently than if they were driven at random can be related to appropriate interpretation of the relevant signs. Large quantitative changes can be made, and the collision-avoidance phenomenon remains. This phenomenon, in a more general context, makes biology `a different game' to ordinary physics.

Signs play an important role in advanced activities through the way complicated signs open up new possibilities, the power of natural language providing a simple illustration of this fact.

The question now arises how semiotic processes manifest and develop, and whether this can happen in the primoridal context which we imagine to be the source of universes and physical laws.  A more global perspective is required, and we now discuss this in the light of Yardley's Circular Theory.

\section{Application of Circular Theory}

Circular Theory\cite{circular} is a work in progress, aimed at expressing structure and function in biological systems in its most basic conceptual form, the key elements being units (`circles'), links between units, and the tendency for units to form (unitisation).

We first discuss the terms unit and link.  Unit is not defined in rigorous terms, the existence of units being something that is discovered though attempts to characterise systems of interest; a unit is something that it is convenient to treat as a whole.  The concept of a unit may usefully be extended to refer to classes that it is convenient to deal with in an analysis, and it may equally well be applied to processes. 

Turning to the concept of link, what is crucial in circular theory is the way systems are able to work together, acting effectively as a single system.  A simple example is provided by a thermostatically controlled system, where a controller, together with a controlled system whose temperature is subject to variation from external inputs, become a system with approximately fixed temperature, while a more complicated case consists in a function present in a computer as a part of a program, interacting with some other system so as to exercise that function.  A server-client situation such as a web browser interacting with a web server illustrates on the other hand a situation of mutual influence.  The point is that there is a special kind of situation of `systems being attuned to each other' that produces highly coordinated behaviour, and this is very relevant to mechanisms and to life generally.  Yet another example is the correlation between the two strands of DNA, in which case the correlations are put to work in the service of copying information.

Intuitively (no attempt will be made here to formulate the concepts rigorously), the point is that the coupling between the systems concerned reduces the range of variation available to the joint system, while still making degrees of freedom available. Arguably, this will tend to happen spontaneously under certain circumstances (as when two clocks are coupled by placing them on a common platform).  This coordination may also be induced by a third influence, as happens during learning involving the development of coordination between two processes.

\subsection{A packing model}

The concepts of circular theory, including the `attunement' concept, can be underpinned by an idea to the effect that what is involved at root is the \textit{packing together} of a set of dynamical systems subject to certain constraints; indeed learning involves the attempt to make systems that are interacting generate activity that conforms to particular constraints.  As an implementation mechanism, we suppose that in place of fixed structures we are concerned in each case with a \textit{collection} of structures distinguished from each other by a set of bits, which are adjusted bit by bit until a high degree of conformance to the relevant constraints is achieved.  This process is equivalent to that of Ross Ashby's ultrastability\cite{ultrastability}.

We can take the idea further by invoking an additional system that can pack other structures together `intelligently', that is to say by recognising signs and responding appropriately, in the manner of semiotic theory.  Such a grouping of three systems can be expected to cohere together more effectively than with situations where there is no such intelligent response to signs.  With such a grouping there is no essential difference between the three components, and all three can be considered interpretants, each interpreting signs originating in the other two systems, and also the interactions between these two.

Conversely, the splitting of a unit into three subunits brings into existence a triadic situation of the kind discussed by Peirce.  What remains when systems disperse in this way is the potential to bond with systems similar to those with which they have previously formed the capacity to bond. In this way we can understand creative development, where new structures form, with new capacities.

These points can be illustrated with analogies from chemistry: (i) if a molecule A can split into two specific molecules B and C, then in a different environment B and C can combine again to form A; (ii) in an extension of the idea, we consider A splitting into three consitutents B, C and D.  In the context of recombination, D can act as a catalyst holding B and C in the correct configuration to enable all three to bond together; (iii) the point about bonding of similar systems is illustrated by the way that if one halogen can bond in a particular place in a specific molecule then a different halogen is likely also to be able to bond in the same place.

\section{Universality, fractality and `turtles all the way down'}

Two complementary forms of change to be considered in the above picture are (i) systems joining together to form one unit, and (ii) a system splitting into a number of units.  This leads to the possibility of a fractal, or scale-free, situation where similar structures exist at all scales.  In this context, some signs would have a universal significance at all levels.  However, as systems become more complex, differentiation and specialisation start to occur.

If the multiple scale picture is correct, we would have a situation where details are governed by finer details which are governed by finer details and so on \textit{ad infinitum}, in conformity with the `turtles all the way down' concept \cite{turtles}.

\section{Cognitive and cultural development}

We first recall what the purpose of the discussion of semiosis and the circular theory has been.  The idea was to be able to treat universe generation as, in essence, a kind of technological development.  The familiar technological development is a product of human beings and brains, and clearly cannot be used to account for universe generation, but our discussion of development in terms of semiosis and circular theory indicates that something analogous to cognitive development (including cultural development, assuming that cognitive development, in a social system, provides a basis for cultural emergence) can occur in a wider context, including that of our postulated primordial system.

The hypothesis then is that primordial constructs of various levels of complexity can form, whose links with other systems including their environment can be equated with `knowing'.  What might such systems come to know?  If their culture acts on the basis of perceived benefit only (as is tending to become the norm in our modern society), then such developments may have limited outcomes.  If wider explorations are not excluded, then developments such as mathematics are possible, which might then be applied to such scientific knowledge as might be discoverable, and subsequently in technological applications including, it is hypothesised, mathematically governed universes that could be beneficial to life.

\subsection{Outliers}

In this connection, Yardley (private communication) notes that an important role in determining the general direction of development is played by \textit{outliers}, that is to say situations encountered that have not yet made effective links with existing structures.  Chance contacts may cause new structures to be built, which structures may on occasion be applicable in a wide range of situations, leading to more extended developments. 
 
\subsection{Mathematical precision}

One important issue is how mathematical precision emerges from a system that is initially very imprecise.  We can usefully consider in this connection Euclidean geometry, a mathematically precise system that emerged through the consideration of properties of the world that were not known with any great precision.  Geometry, like any mathematical enterprise, is a symbolic activity that does not depend in any essential way on interaction with the world.  It was, nevertheless, inspired by knowledge of real point-like objects and approximate straight lines.  By retreating into symbolism one escapes inconvenient facts about the world and is able to create a system that has a certain resemblance to the world even though there is no exact correspondence.  The Euclidean plane, is in essence, a fantasy that one can address through symbols even though the real world does not correspond exactly to it.  However, in this case the correspondences between the Euclidean world and the real world are sufficiently close that Euclidean geometry is of value in the real world, but this is something that has to be discovered through observation rather than taken for granted.
  
\subsection{Generation of space and physical universes}

In our ordinary world, Euclidean geometry is simply a system that provides a good model for phenomena in space, using specialised techniques to connect the model with the reality.  From the perspective of our primordial community, it conversely provides a model for \textit{forming} a universe-system (more generally, physical laws provide a basis for forming the corresponding physical reality). The model is not the technology, any more than understanding the sphere equates to the existence of \textit{physical} spheres.  We hypothesise however that some such technology, which in due course we may ourselves be able to understand, was discovered at the primordial level, and forms the basis upon which physical universes are generated.  Mathematical precision exists only in the world of discourse, and is realised to whatever degree is possible by technology.

Symmetry and symmetry breaking may play a key role here, in view of the fact that conceptually symmetry is defined in terms of transformations that may have physical correlates, while at the same time symmetry is found to play an important role in actual physics.

In this picture locality is understood as an \textit{emergent} property, analogous to the frequency of a physical process.  Just as in some circumstances frequencies of physical processes become well defined, with different frequencies becoming independent of each other as far as linkages are concerned, in this case location becomes a well defined quantity, with different locations becoming independent of each other.  Quantum entanglement and wholeness, on the other hand, would be derivative of the units of circular theory.  More generally, the high degree of correlation associated with the packing model can be expected to be manifested in phenomena similar to those associated with quantum mechanics.
  
\section{Discussion}

We have addressed in a natural way Wheeler's question of how observer-participation can lead to the emergence of specific laws of nature in particular systems.  The key point is the fact that \textit{the interpretation of signs changes the game}, facilitating the emergence of new kinds of system and process, which are correlates of cognitive and cultural development that, in the present context, lead to emergent laws.  In this picture, the responsible system or systems are the determiners of the observed laws, rather than the laws concerned being presumed absolute, or derivable from some mathematical analysis.

One can imagine a scenario whereby conventional science would be forced similarly to renounce the idea of a Final Theory.  We already have a situation where some theory X (e.g. the Standard Model) proves inadequate and theory Y (e.g. string theory) is proposed to take its place.  Then certain further issues lead to the idea that the real `fundamental theory' is Z (e.g. M-theory).  At each stage, however, the supposed fundamental theory gets farther from what is accessible by experiment, and its connections with reality become more obscure.

The idea that nature at some deeper level has biological aspects is not fundamentally absurd, and has been previously explored by authors such as  Smolin\cite{Smolin} and Pattee\cite{Pattee}.  The above analysis has explored some aspects of the `biological logic' applicable to such a scenario, in particular the mechanics of development, which could lead to what might be termed `extended mind'.  Faculties such as mathematical intuition, difficult to account for in conventional ways, might be manifestations of the extended mind, which might also be related to experiences of meaning in art.

To what extent can these proposals be considered scientific in character?  While the absence of a fixed, universal mathematical law may seem at first sight to be a radical departure from scientific tradition, the idea that the laws manifested in the laboratory are emergent rather than fundamental is already a feature of string theory.  And, as practiced, biology is a science that makes extensive use of  phenomenology (e.g. that of chemical reactions), and concepts specific to biology, and typically makes less use of the methods of theoretical physics (i.e. mathematical models).

A typical biological concept is the idea that particular systems (e.g. the immune system) have particular functions.  Such concepts have value in interpreting what one finds and in guiding investigations. The ideas expounded here can be expected to be of similar value in constructing models where conventional methods prove inadequate.

Some scientists have accepted the idea that not everything can be characterised in quantitative terms, asserting however that the only real knowledge is that based on scientific measurement; but alternatives \cite{INBIOSA, shift}, offering a broader understanding of what constitutes knowledge, are possible.  The present discussion offers some insight into what is involved in that latter position.  Nature is pervaded by patterns (signs) which through practice we have become expert in interpreting, a process that has pragmatic value even if it is not amenable to the traditional quantitative methodology.  If the picture developed here is correct, there is much more in the way of meaning to be found in the natural world by such means than can be found through the traditional methodology of science.
  
\subsection{Acknowledgments}
I am grateful to Ilexa Yardley for numerous discussions that have helped shape these ideas, and to Dr.\ Plamen Simeonov for helpful comments on the manuscript.

\end{document}